\newcount\equnumber
\def\eq(#1){\global\advance\equnumber by 1 
    \expandafter\xdef\csname !#1\endcsname{\the\equnumber}
    \eqno(\the\equnumber)}                               
\def\(#1){(\csname !#1\endcsname)}
\def\fr#1/#2{{\textstyle{#1\over#2}}} 
\magnification=1200      		 
\font\title = cmbx10 scaled 1440        
\def\h{\fr1/2}
\def\la{\lambda}
\magnification = \magstep1
\baselineskip = 15pt
\parskip=3pt
\def\head#1{\bigskip\noindent{\bf #1}}

\def\caption#1#2{
{\narrower

\baselineskip = 12pt
\medskip
\noindent {\bf Figure #1.} #2

}}

\input epsf

\centerline{\title  From Einstein's 1905 Postulates} 
\medskip
\centerline{\title  to the Geometry of Flat Space-Time}
 
\bigskip
\centerline{N. David Mermin}
\centerline{Laboratory of Atomic and Solid State Physics}
\centerline{Cornell University, Ithaca, NY 14853-2501}

\bigskip

{

\narrower
\baselineskip = 12pt

Minkowski diagrams in 1+1 dimensional flat space-time are given a
strictly geometric derivation, directly from two gedanken experiments
incorporating the principle of the constancy of the velocity of light
and the principle of (special) relativity.  Rectangles of photon
trajectories play a central role in determining the simultaneity
convention and in establishing the invariance of the interval.

}

\bigskip

For the hundredth anniversary of {\it Zur Elektrodynamik bewegter
K\"orper\/} I would like to describe a derivation of the geometry of
flat Minkowski space-time straight from Einstein's two {\it
Voraussetzungen\/}.  No use will be made of either the Lorentz
transformation or the invariant interval, which emerge as secondary
features of the special-relativistic space-time geometry that follows
directly from applying the two postulates to two {\it gedanken\/}
experiments.  I consider only one spatial dimension, commenting at the
end on the higher dimensional generalization.

Assisting me in this approach to space-time geometry will be Alice and
Bob, who have appeared in expositions of cryptography for many years.
With the development of quantum cryptography and quantum information
theory they have become well known to many physicists during the past
decade.  They can also be invaluable in making expositions of
relativity more readable and concise.

\head{1. Alice's diagram: equilocs, equitemps, and scale factors}

Alice, who uses the terminology appropriate to an inertial frame of
reference, represents events as points in a plane diagram.  In what
follows I shall use ``event'' to signify the representative
point as well as the event itself.  Events that happen at the same
place in Alice's frame are all placed on the same straight line: an
{\it equiloc\/}.  (I would have preferred the term {\it isotop\/} but
the similarity --- not to mention the identity in German --- to
``isotope'' makes this unacceptable.)  Equilocs associated with
different places must be parallel, for otherwise they would intersect
and the point of intersection would represent a single event happening
in two different places.

The distance between equilocs in Alice's diagram is proportional to
the distance between the places they represent.  The proportionality
constant $\lambda$ (or $\lambda_A$ if we wish to emphasize its use by
Alice) specifies, for example, the number of centimeters of diagram
separating equilocs associated with events one foot apart in space.
Her unit of distance, the {\it foot\/} (abbreviated $f$), is defined
to be a light nanosecond: 1 f = 0.299792458 m.  Although this term
competes with the name still used in a few backward countries for a
distance of 0.3048 m, Alice's foot falls short of the Engish foot by a
mere 1.6\%.  What makes the word irresistible is its phonetic
resemblance to $\phi\omega\tau o \varsigma$ (light).

Alice positions events along her equilocs so that events she takes to
be simultaneous also lie along a straight line: an {\it equitemp\/}.
(The more beautiful {\it isochron\/} is linguistically incompatible
with ``equiloc'' and, like ``isotop'', is also used for other
scientific purposes.)  Equitemps associated with different times must
also be parallel, since an event cannot happen at two distinct times.
Alice's equitemps cannot be parallel to her equilocs, since specifying
both a place and a time specifies a unique event, and therefore a
unique point in her diagram, but aside from that, she can choose the
angle $\Theta$ (more fully, $\Theta_ A$) between her equitemps and
equilocs as she wishes.

The distance between equitemps in Alice's diagram is proportional to
the time between the events they represent. It is convenient for her
to use the same scale factor $\lambda$ for equitemps associated with
events 1 ns apart, as she uses for equilocs associated with events 1 f
apart.  It is also convenient to introduce a second scale factor $\mu$
which specifies the distance along a given equiloc of events whose
times differ by 1 ns (or the distance along a given equitemp of events
whose locations differ by 1 f).  Evidently (see Figure 1) the two
scale factors are related by $$\lambda = \mu\sin\Theta.\eq(Theta)$$

\midinsert
\epsfxsize=4.5 truein
\centerline{\epsfbox{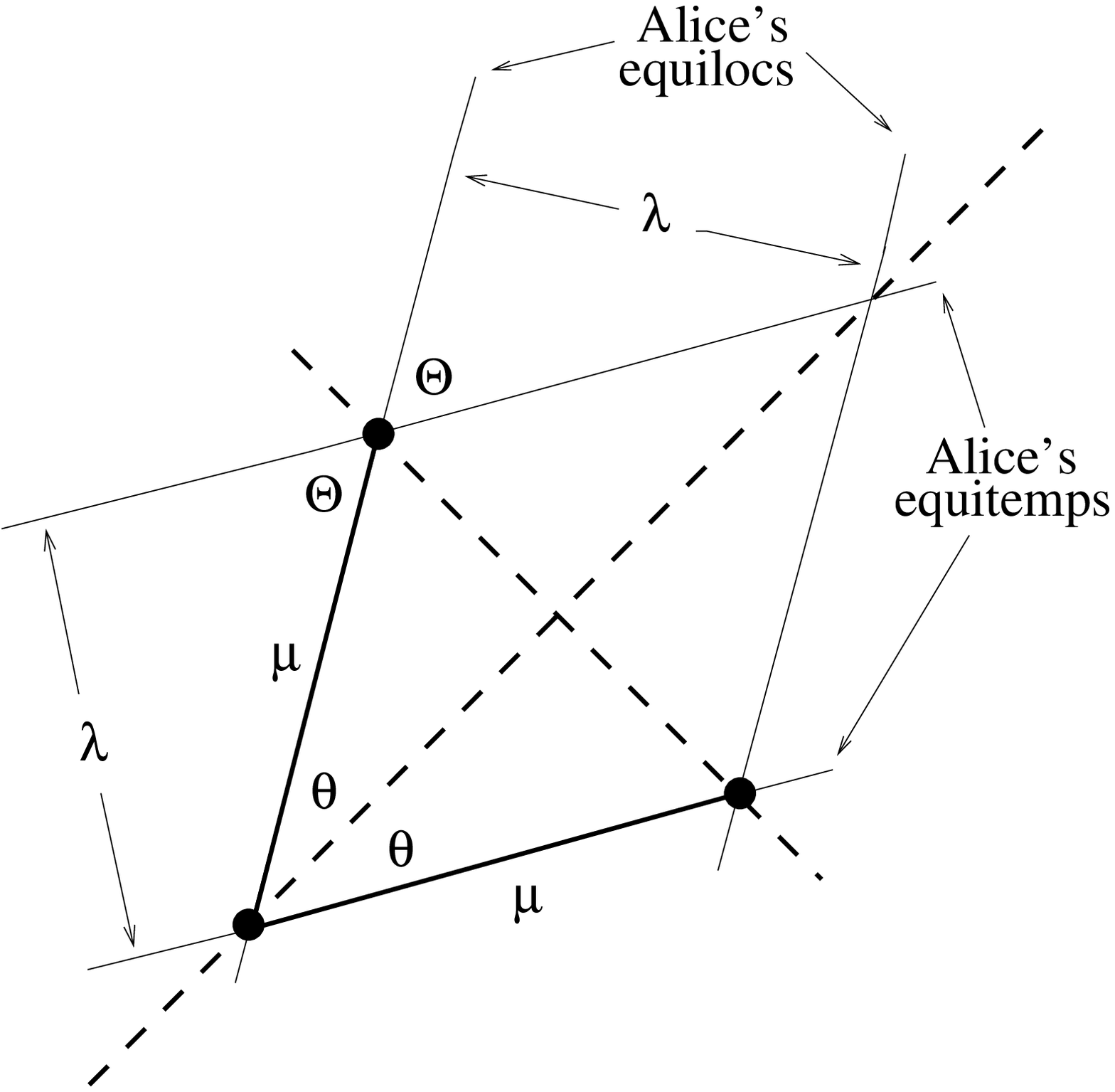}}

\caption1{The parallel lines tilting slightly upward
to the right are equitemps.  Each represents events that happen at the
same time in Alice's frame.  The times associated with the two
equitemps are 1 ns apart.  The parallel lines tilting steeply upward
to the right are equilocs.  Each represents events that happen in the
same place.  The positions associated with the two equilocs are 1 f
apart.  Alice's scale factor $\lambda$ is the distance in the diagram
between the equilocs or between the equitemps.  Her scale factor $\mu$
is the length in the diagram of the (more heavily drawn) segments of
the equitemps and equilocs between points (black circles) that
represent events 1 f or 1 ns apart.  Evidently $\lambda =
\mu\sin\Theta$ where $\Theta = 2\theta$ is the angle between equilocs
and equitemps.  The two equitemps and two equilocs bound a rhombus of
area $\la\mu$.  Both diagonals of the rhombus are photon space-time
trajectories, since they connect events 1f and 1ns apart.  Being
diagonals of a rhombus they are perpendicular and bisect the angles at
the vertices.}

\endinsert

Because the speed of light is 1 f/ns and because of the relation Alice
imposes on the scale factors for her equilocs and equitemps, it
follows from Figure 1 that the space-time trajectories of two
oppositely moving photons present at a given event, bisect the angles
between the equiloc and equitemp on which that event lies, and that
trajectories of oppositely moving photons are perpendicular.  

Alice orients her diagram as in Figure 1, so the two families of
photon trajectories make angles of $\pi/4$ with the vertical direction
on the page.  Her equilocs and equitemps are symmetrically disposed on
either side of the diagonal directions of the photon trajectories, at
angles $\theta = \fr1/2\Theta$ (or $\theta' = \fr1/2\Theta'$,\
$\Theta' = \pi - \Theta$) to the photon lines.  Alice further orients
her page so that the equitemps lie below the diagonals (i.e. they are
more horizontal than vertical) and the equilocs lie above them
(i.e. they are more vertical than horizontal).  Finally, she orients
the page so that equitemps higher on the page are associated with
later times.  

The structure and orientation of Alice's diagrams are
then uniquely determined except for two free parameters which we can
take to be her scale factor $\lambda$ specifying how many centimeters
of diagram separate two equitemps associated with events one ns apart,
and her angle $\Theta$ between her families of equilocs and equitemps.

\head{2. Bob's use of Alice's diagram with his own equilocs and
equitemps.}

Bob, who moves uniformly with velocity $v$ in Alice's frame, notes the
same events that she does, and is free to record them in another
diagram of his own making, appropriate to his proper frame.  But since
all events of interest are already present as points in Alice's
diagram, rather than putting events into a new diagram, he can try to
impose on Alice's his own set of equilocs and equitemps.  His equilocs
are entirely straightforward: they are just the space-time
trajectories of objects that move with velocity $v$ in Alice's frame
and are therefore also parallel straight lines, making the appropriate
angle with Alice's equilocs, and, like Alice's, tilted by less than
$\pi/4$ from the vertical, if $v$ is less than 1 f/ns.

\midinsert
\epsfxsize=3.5 truein
\centerline{\epsfbox{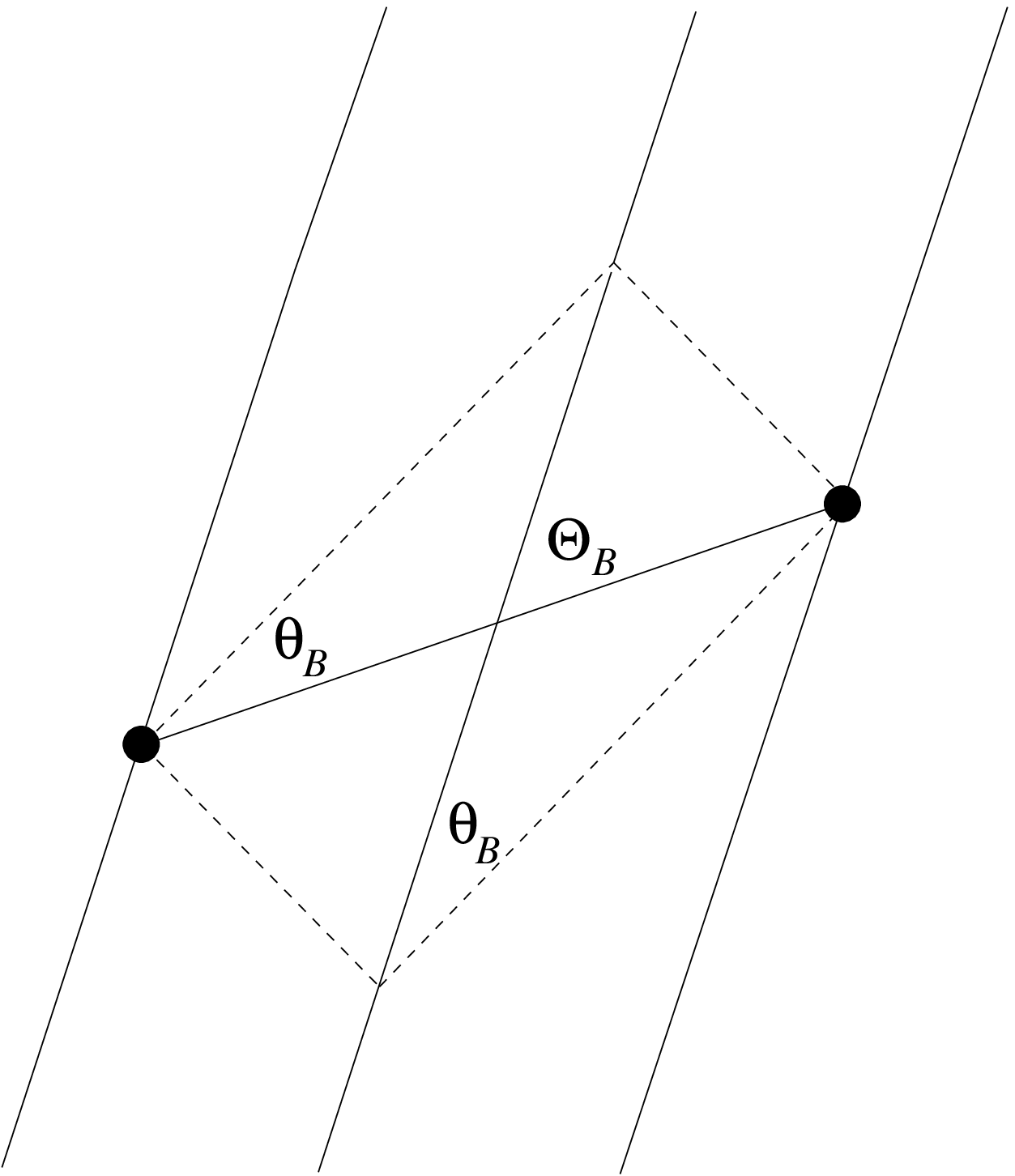}}

\caption2{The three equally spaced parallel lines are
two of Bob's equilocs and a third, associated with a position exactly
halfway between the positions associated with the other two.  Two
photons leave a point on the middle equiloc and travel in opposite
directions to the positions of the two outer equilocs, where they are
reflected back to their starting equiloc.  Because the photons start
midway between the outer positions, their arrival at the outer
positions, events marked by the two black circles, are simultaneous
events in Bob's frame and the line joining them is an
equitemp for Bob.  It is evident from the symmetry of the rectangle
formed by the four photon trajectories that the angle $\theta_B$
between Bob's equitemp and a photon trajectory is the same as the
angle $\theta_B$ between his equilocs and that photon trajectory.} 

\endinsert

To establish the character of Bob's equitemps in Alice's diagram we
must, for the first time, invoke Einstein's postulates.  The velocity
of light is 1 f/ns, independent of the velocity of the source, in both
Bob's frame and Alice's.  Therefore two events determine an equitemp
in Bob's frame if and only if oppositely moving photons emitted at
each event meet midway between their locations, or, equivalently, if
it is possible for oppositely moving photons emitted together midway
between their locations to arrive at the events just as they happen.

Two events satisfying these conditions are pictured as two black
circles in Alice's diagram in Figure 2.  The three parallel lines are
the two equilocs of Bob on which the events lie and his equiloc midway
between them.  The trajectories of four photons, demonstrating that
two events on his outer two equilocs are simultaneous in Bob's frame,
form a rectangle.  The more vertical diagonal of that rectangle is the
middle equiloc of Bob.  The more horizontal diagonal connects the
points representing Bob's two simultaneous events and is therefore an
equitemp in his frame.  It is evident from the symmetry of the
rectangle that these two diagonals are symmetrically disposed about
the photon lines: lines in Alice's diagram connecting two events that
are simultaneous in Bob's frame, make the same angle $\theta_B$ with
the photon lines as Bob's equilocs do.

So Bob's equitemps are straight lines, making an angle $\Theta_B =
2\theta_B$ with his equilocs, symmetrically disposed with his equilocs
about the photon lines, just as Alice's are with angles $\Theta_A =
2\theta_A$.  Given a diagram with both their sets of equitemps and
equilocs there is therefore no way to tell which of them made the
diagram first and which then added to it their own equitemps and
equilocs: the frame independence of the velocity of light is
explicitly consistent with the principle of relativity.

\head{3.  Relation between Alice's and Bob's scale factors: light rectangles.}

It remains is to establish the relation between the scale factors
$\lambda$ (or $\mu$) used by Alice and Bob.  This has a simple
geometric formulation in terms of rectangles of photon trajectories,
like the one in Figure 2.  One can associate with {\it any\/} two
events a unique rectangle whose four sides are segments of the two
pairs of photon lines passing through each of the events, as shown in
Figure 3.

\midinsert
\epsfxsize=1.5 truein
\centerline{\epsfbox{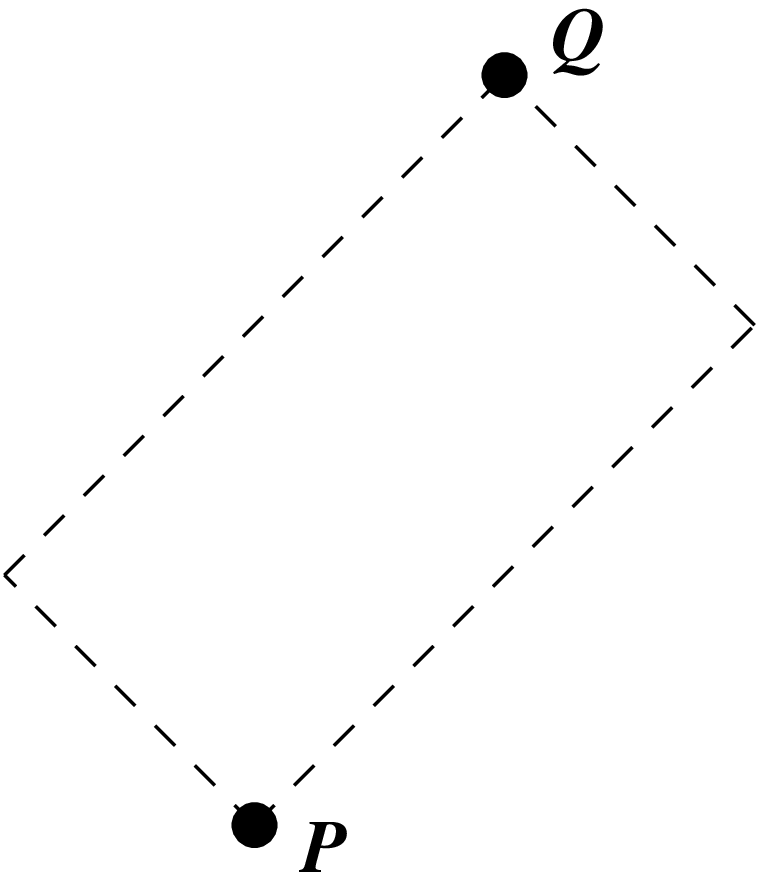}}

\caption3{Two events $P$ and $Q$ determine a unique rectangle of
photon line segments --- a {\it light rectangle\/} ---with the events
at diagonally opposite vertices.}

\endinsert

The relation between Alice's and Bob's scale factors is determined by
the fact that the area of such a {\it light rectangle\/} for two
events on an equiloc of Alice, a time $T$ apart in Alice's frame, must
be the same as the area of the light rectangle for two events on an
equiloc of Bob, the same time $T$ apart in Bob's frame.This equality
of areas follows directly from the physical fact that if Alice and Bob
each {\it looks\/} at a clock carried by the other, the rate at which
each {\it sees\/} the other's clock running will differ from the rate
of their own clock by the same factor $f$, as required by the
principle of relativity and the source independence of the speed of
light.  That this reciprocity of Doppler shifts has this immediate
geometric consequence is demonstrated in Figure 4.

\midinsert
\epsfxsize=4 truein
\centerline{\epsfbox{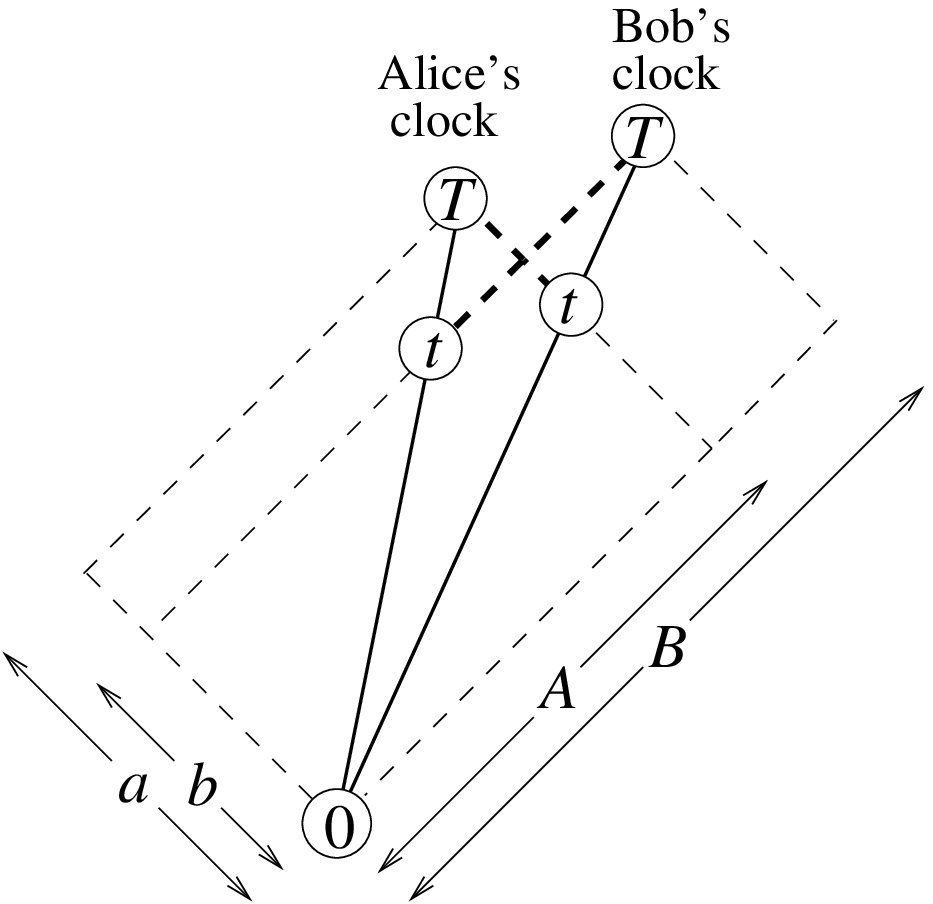}}

\caption4{The space-time trajectories of Alice's
clock and Bob's intersect when both read 0.  The heavier dashed line
segments are photon trajectories demonstrating that when the clock of
each reads $T$, each sees the clock of the other reading $t$.  Two
light rectangles, one with sides $A$ and $a$, the other with sides $B$
and $b$ are formed by the photon trajectories through the two clocks
reading $T$ and the two (coincident) clocks reading 0.}

\endinsert

When Alice and Bob are together they set their clocks to 0.  Then they
move uniformly apart.  When their clocks read $T$ each looks back at
the other's clock and sees it reading $t$.  Figure 4 reveals that the
ratio $t/T$ associated with Alice's clock is equal to $b/a$, while the
ratio $t/T$ associated with Bob's is equal to $A/B$.  Therefore $b/a =
A/B$, so $Aa = Bb$.  But $Aa$ is the area of the light rectangle with
Alice's clock reading $T$ and 0 at opposite vertices, while $Bb$ is
the area of the corresponding light rectangle for Bob's clock.
 
Figure 5 shows that the area of such a light rectangle is $$\Omega =
\h\la\mu T^2, \eq(lightarea)$$ so the simplest analytical expression
of the geometric connection between Alice's and Bob's scale factors is
just that the product $\la\mu$ of the two scale factors is independent
of frame: $$\la_A\mu_A = \la_B\mu_B.\eq(lamu)$$

\midinsert
\epsfxsize=3.5 truein
\centerline{\epsfbox{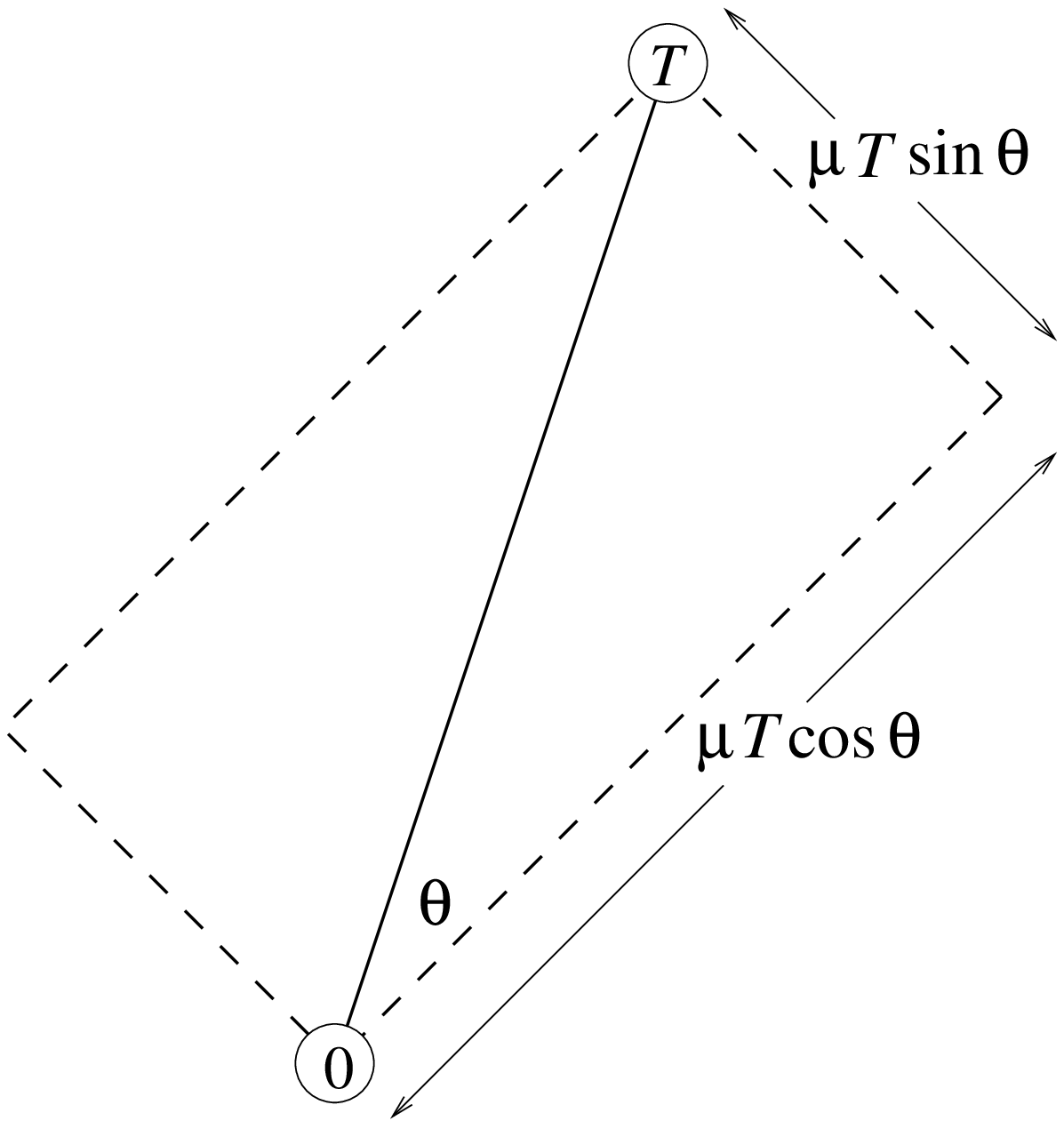}}
\caption5{The length of an
equiloc between two readings of the same clock a time $T$ apart is
$\mu T$, where $\mu$ is the scale factor for the equiloc.  The area of
the light rectangle with the events at opposite vertices is $\Omega = (\mu
T)^2\sin\theta\cos\theta$ $= \fr1/2\mu^2\sin\Theta\, T^2$ $= \h\mu\la
T^2$.}

\endinsert

It is useful to have an expression for this frame-independent product
of scale factors, which we take in the form $$\Omega_0 =
\h\la\mu.\eq(Omega0)$$ If $\la$ and $\mu$ are given in cm of diagram
per ns of time (or f of space), then $\Omega_0$ has dimensions of
cm$^2$/ns$^2$ or cm$^2$/f$^2$. It is convenient to define for any
frame --- say Bob's --- a {\it unit light rectangle\/} as one with a
diagonal that is an equiloc for Bob connecting events 1 ns apart.
Figure 2 establishes that the other diagonal is an equitemp for Bob,
and the relation between scales on equitemps and equilocs establishes
that the vertices on the equitemp are events 1 f apart in his frame.
Unit light rectangles in all frames have the same frame-independent
area, $\Omega_0 = \h\lambda\mu.$

\head{4. Light rectangles and the invariant interval.}

Abstracting from Alice and Bob, we can say that the area (in units of
$\Omega_0$) of the light rectangle determined by two time-like
separated events is the square of the time between them in the frame
in which they happen in the same place.  This is precisely the
definition of the {\it squared interval\/} $I^2$ between the events,
so $$I^2 = \Omega/\Omega_0.\eq(interval)$$

Because the same scale factor $\la$ is associated with equilocs and
equitemps this also holds for the interval between two
space-like separated events: $I^2 = \Omega/\Omega_0$, where $I^2$ is
the square of the distance between the events in the frame in which
they happen at the same time, and $\Omega$ is the area of their light
rectangle.  

The light rectangle for two light-like separated events degenerates to
a line, which has zero area, again agreeing with the definition of the
interval.

\head{5. Interval in terms of coordinates.}

Figure 6 relates the geometric representation of the interval as a
light rectangle to its expression in terms of coordinates in a
particular frame, showing that the squared interval between two
time-like separated events is $I^2 = T^2-D^2$ where $T$ and $D$ are
the time and distance between the events in any frame.

\midinsert
\epsfxsize=4 truein
\centerline{\epsfbox{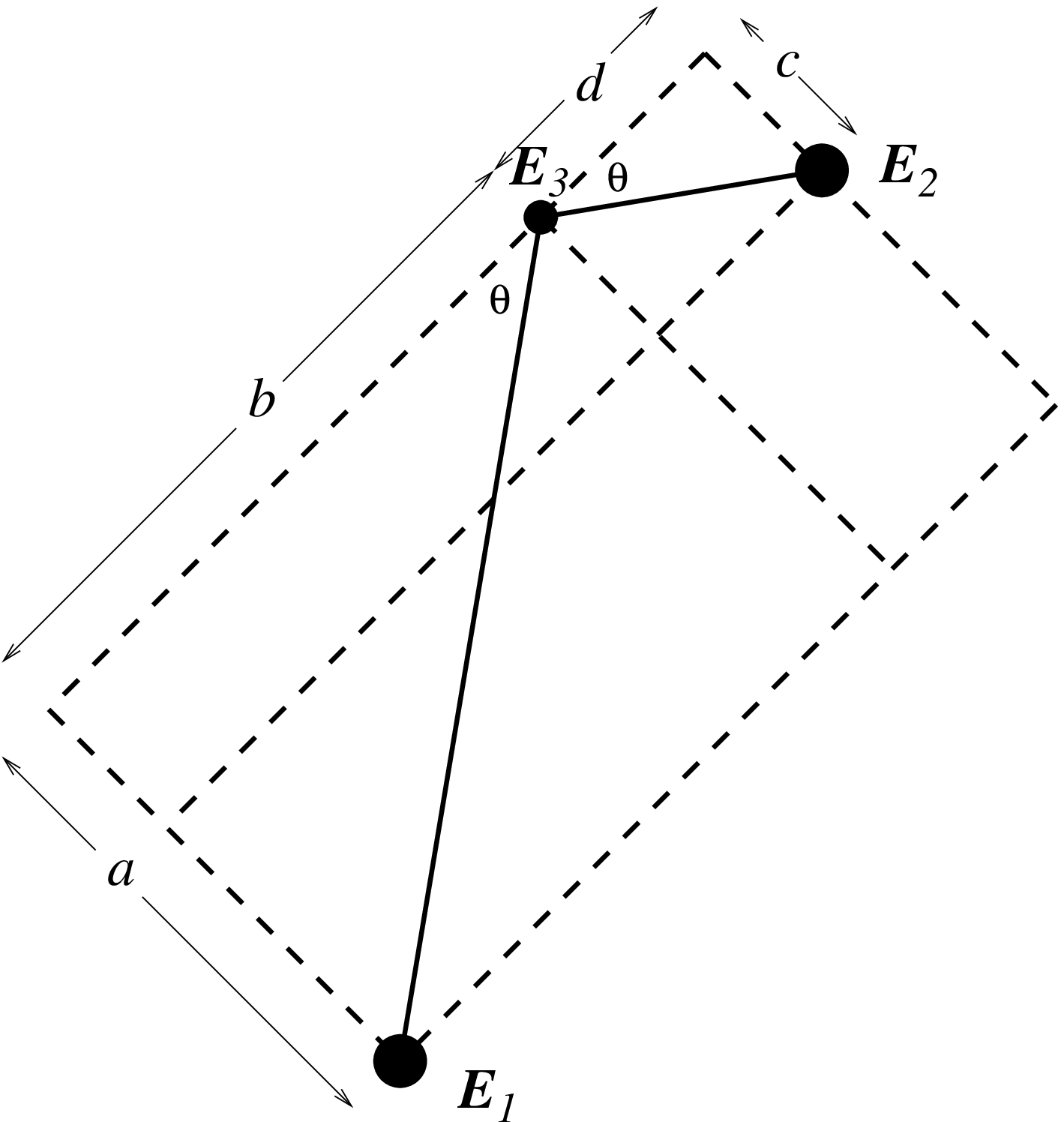}}

\caption6{The two larger black circles are two time-like separated
events $E_1$ and $E_2$.  The dashed lines are photon trajectories.
The two solid lines are Alice's equitemp and equiloc, which intersect
at an event $E_3$.  The squared distance between $E_1$ and $E_2$ in
Alice's frame is proportional to the area of the rectangle with $E_3$
and $E_2$ at opposite vertices.  Her squared time between $E_1$ and
$E_2$ is proportional to the area of the rectangle with $E_1$ and
$E_3$ at opposite vertices.  The squared interval between $E_2$ and
$E_1$ is proportional to the area of the rectangle with $E_1$ and
$E_2$ at opposite vertices.}

\endinsert

The two large black circles are the events $E_1$ and $E_2$.  The solid
lines are Alice's equitemp and equiloc, which intersect at an event
$E_3$, where they make the same angle $\theta$ with a photon
trajectory.  As a result, the right triangle with sides $c$ and $d$ is
similar to the right triangle with sides $a$ and $b$, so $a/b = c/d$
and therefore $$ad = bc.\eq(=rects)$$ The squared interval between
$E_1$ and $E_2$ is proportional to the area $(a-c)(b+d)$ of their
light rectangle: $$ I^2 = (a-c)(b+d)/\Omega_0,\eq(e12)$$ which
\(=rects) simplifies to $$ I^2 = (ab-cd)/\Omega_0.\eq(I12)$$

\noindent But $ab/\Omega_0$ is the squared interval between $E_1$ and
$E_3$ while $cd/\Omega_0$ is the squared interval between $E_2$ and
$E_3$.  Since $E_1$ and $E_3$ are on an equiloc in Alice's frame, the
squared interval between them is $T^2$, the square of Alice's time
between them; since $E_2$ and $E_3$ are on an equitemp in Alice's
frame the squared interval between them is $D^2$, the square of
Alice's distance between them.  But since $E_3$ happens at the same
place as $E_1$ and the same time as $E_2$ in Alice's frame, $T$ and
$D$ are also Alice's time and distance between $E_1$ and $E_2$.  So
$$I^2 = T^2 - D^2.\eq(coordint)$$ The analogous argument for
space-like separated events follows from reflecting Figure 6 in any of
the photon trajectories.

\midinsert
\epsfxsize=4.5 truein
\centerline{\epsfbox{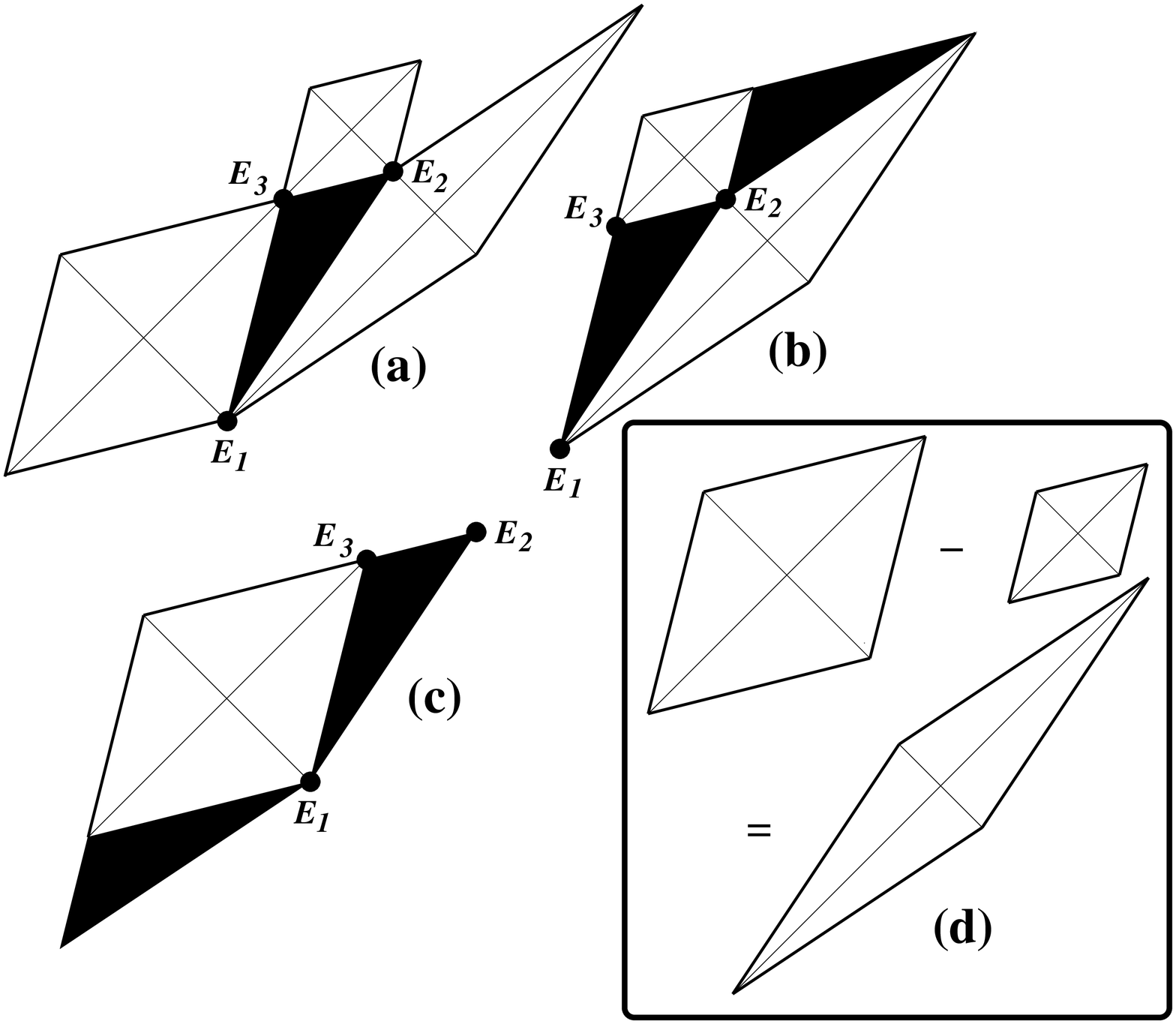}}

\caption7{A proof of $I^2 = T^2 - D^2$ 
more purely geometric than that of Figure 6.  Each of the three light
rectangles has been replaced in part (a) by a rhombus of twice the
area.  The quadrilaterals bounding parts (b) and (c) are identical,
leading to the relation between areas shown symbolically in part (d).}

\endinsert

A purely geometric proof of \(coordint) that avoids even the
tiny bit of analysis in \(=rects)-\(I12) is given in Figure 7.  
Part (a) of Figure 7 reproduces the content of Figure 6, except that
the three light rectangles associated with the three pairs of events
have been replaced by three rhombi, each with twice the area of the
rectangle it replaces.  In part (b) the two smaller rhombi have been
combined with two copies of the triangle formed by the three events to
form a quadrilateral.  In part (c) the largest rhombus is combined
with two copies of that triangle to form that same quadrilateral,
thereby demonstrating that the area of the largest rhombus is the sum
of the areas of the two smaller ones, leading to the relation between
areas shown in part (d).

\midinsert
\epsfxsize=3.5 truein
\centerline{\epsfbox{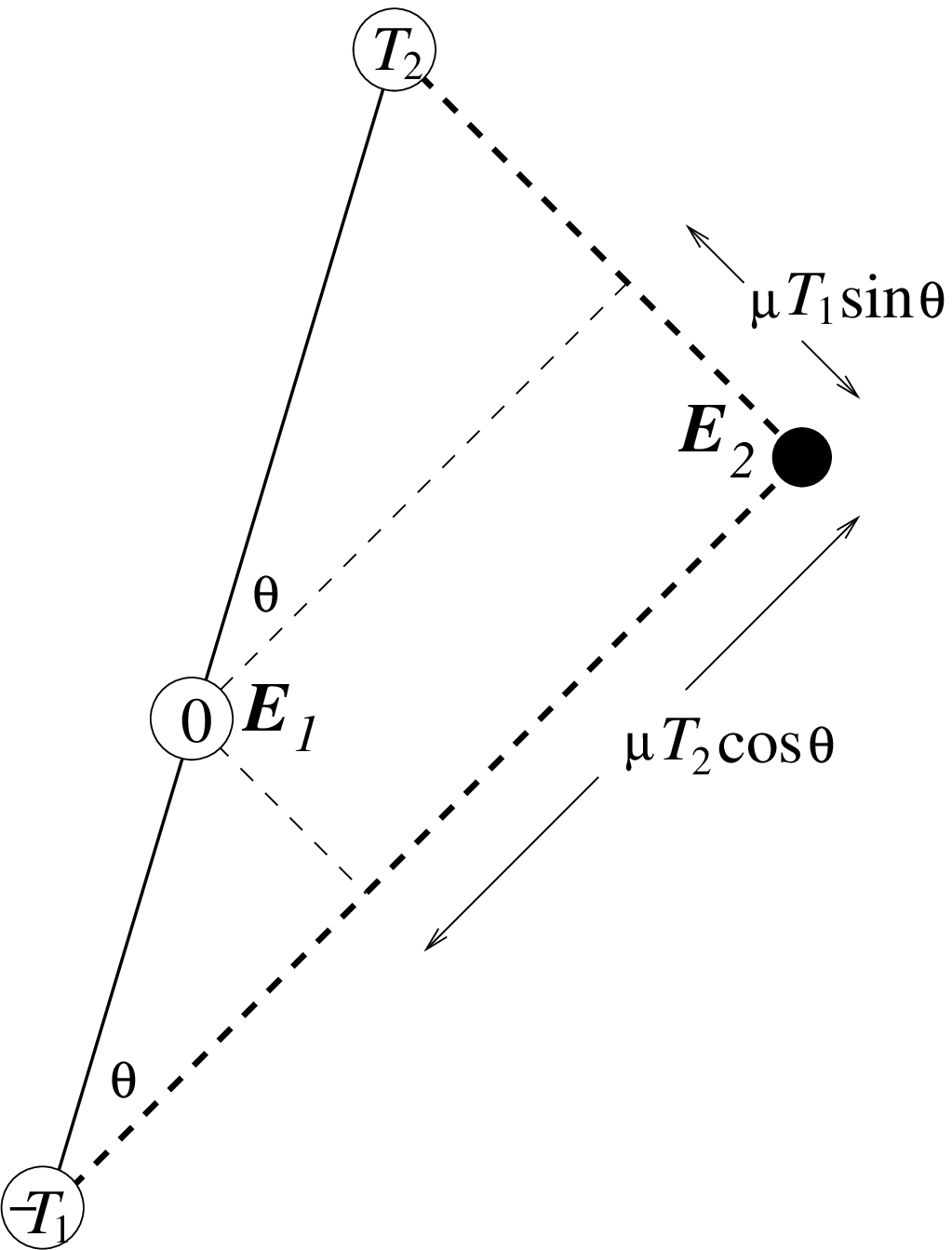}}

\caption8{The solid line is
Alice's equiloc.  Three readings of her clock are shown.  It is seen
at event $E_2$ to read $-T_1$; it reads 0 at event $E_1$; and it reads
$T_2$ when $E_2$ is seen at the location of the clock.  The area of
the light rectangle with $E_1$ and $E_2$ at opposite vertices is
$(\mu T_1 \sin\theta)(\mu T_2\cos\theta) = \Omega_0T_1T_2$.}

\endinsert

\head{6. Measuring the interval with a single clock; an application of
light rectangles.}

There is an elegant way to measure the interval using only light
signals and a single clock [1].  Figure 8 illustrates the method for two
space-like separated events and demonstrates why it works.  The events
are $E_1$ and $E_2$.  A uniformly moving clock (stationary in Alice's
frame) is present at $E_1$.  Call the reading of the clock when $E_1$
takes place 0.  When $E_2$ takes place the clock is seen at $E_2$ to
read $-T_1$.  And when $E_2$ is seen to take place at the clock, the
clock reads $T_2$.  The segments of Alice's equiloc between her clock
reading 0 and reading $T_1$ or $T_2$ have length $\mu T_1$ or $\mu
T_2$ where $\mu$ is Alice's scale factor. Consequently the area of the
light rectangle with $E_1$ and $E_2$ at opposite vertices is

$$\Omega = (\mu
T_1 \sin\theta)(\mu T_2\cos\theta) = \h\mu^2\sin\Theta\, T_1T_2
=\h\mu\la T_1 T_2 = \Omega_0 T_1 T_2.\eq(seeint)$$ The squared
interval $I^2$ between $E_1$ and $E_2$ is $\Omega/\Omega_0$
and therefore $$I^2 = T_1 T_2.\eq(seeint1)$$

In much the same way, Figure 9 establishes for time-like separated
events the relation \(seeint1) between the interval and the three
readings on Alice's clock.

\midinsert
\epsfxsize=4 truein
\centerline{\epsfbox{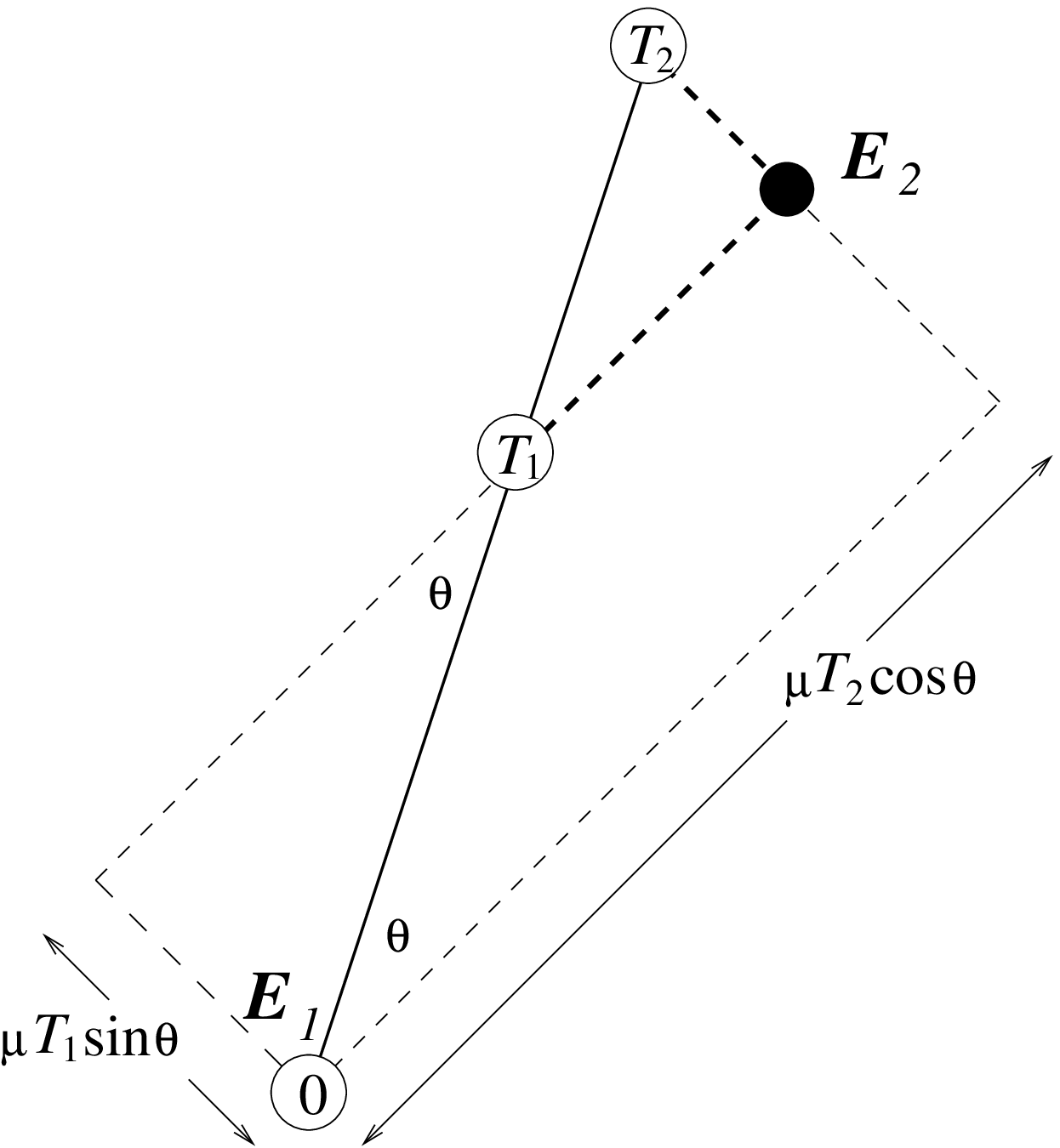}}

\caption9{The solid line is Alice's equiloc.  Three readings of her
clock are shown.  It reads 0 at event $E_1$; it is seen at event $E_2$
to read $T_1$; and it reads $T_2$ when $E_2$ is seen at the location
of the clock.  The area of the light rectangle with $E_1$ and
$E_2$ at opposite vertices is $(\mu T_1 \sin\theta)(\mu T_2\cos\theta)
= \Omega_0T_1T_2$.}

\endinsert
\head{7. Shapes of light rectangles and relative velocities of
frames.}  What is the relation between the {\it shapes\/}
of light rectangles whose diagonals are equilocs in different frames?
Since the shape for any given frame is independent of the size, it
suffices to consider two light rectangles in Figure 10.  One has as
its diagonal Bob's equiloc between events $P$ and $R$, and the other,
Alice's equiloc between events $P$ and $Q$.  The event $Q$ has been
chosen so that the lighter solid line between $Q$ and $R$ is an
equitemp in Alice's frame.  The aspect ratios of Bob's and Alice's
light rectangles are $b/B$ and $a/A$.

\midinsert
\epsfxsize=4.5 truein
\centerline{\epsfbox{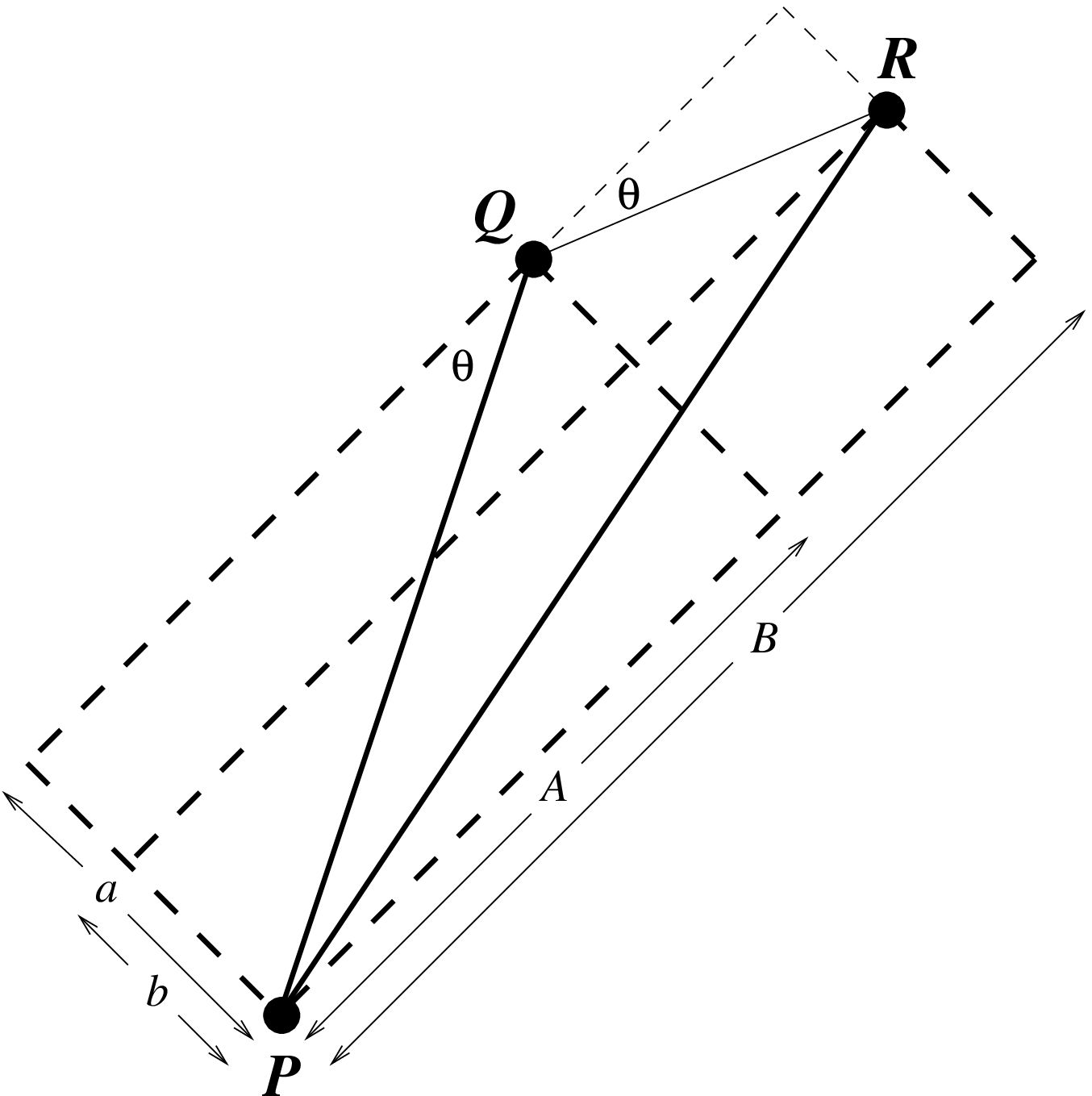}}

\caption{10}{The line from $P$ to $R$ is an equiloc in
Bob's frame.  The lines from $P$ to $Q$ and from $Q$ to $R$, making the
same angle $\theta$ with the photon line through $Q$, are an equiloc
and equitemp in Alice's frame.  Bob's velocity $v$ in Alice's frame is
the ratio of the lengths of these two lines.}

\endinsert
   
We can relate these two aspect ratios to Bob's velocity $v$ in Alice's
frame, by noting that $v$ is the ratio of the length of Alice's
equitemp, from $Q$ to $R$, to the length of her equiloc, from $P$ to
$Q$.  Each of these two lines is the hypothenuse of a right triangle
whose other sides are photon trajectories, and because the lines are
an equitemp and equiloc for Alice, the two right triangles are
similar.  Consequently the ratio $v$ of the hypothenuse lengths is
equal to the ratio of the lengths of either pair of corresponding
sides, and we have $$v = {a-b \over a} = {B-A \over A}. \eq(vratio)$$
These relations tell us that $$b/a = 1-v\ \ {\rm and}\ \ B/A =
1+v.\eq(intermed)$$ Consequently the aspect ratio $b/B$ of Bob's light
rectangle is related to the aspect ratio $a/A$ of Alice's by $$ {b/B
\over a/A} = {1-v \over 1+v}.\eq(aspectratio)$$

With this information we can extract the quantitative expression for
the relativistic Doppler shift from Figure 4.  Applied to the two
photon rectangles in that Figure, \(aspectratio) tells us that $$ {1-v
\over 1+v} = (A/B)(b/a), \eq(aspectratio1)$$ where $v$ is the velocity
of Bob in Alice's frame of reference.  But the equality of the
 rates $f$ at which Alice or Bob sees
time passing on the other's clock, as measured by their own clock,
requires that 
$$A/B = f = b/a,\eq(samearea)$$ and therefore  $$f
= \sqrt{ {1-v \over 1+v} }.\eq(dopp1)$$

\head{8. Doppler shifted lengths; another application of light rectangles.}

The relation \(aspectratio1) also provides a diagrammatic
demonstration of the less frequently noted fact that a train moving
toward one with speed $v$ is {\it seen\/}, compared with stationary
objects, to be {\it longer\/} than its proper length by the Doppler
factor $\sqrt{{1+v \over 1-v}}$.  The white circle in the upper right
of Figure 11 is the event in which the front of the train reaches Bob.
The second  white circle in the lower left, connected to the first
circle by the heavy photon line is the event at the rear of the train
that Bob sees when the front of the train reaches him.

\midinsert
\epsfxsize=5 truein
\centerline{\epsfbox{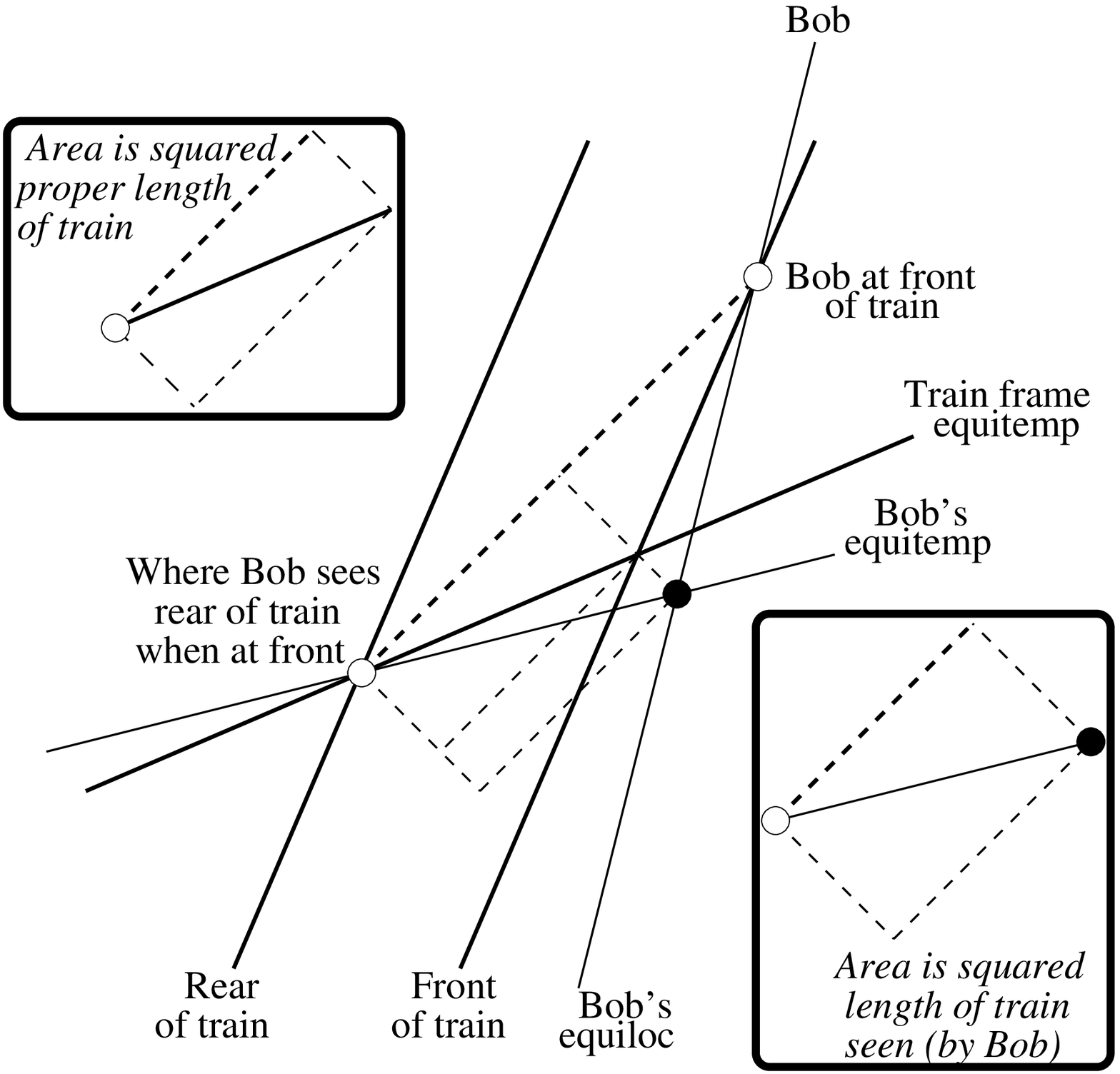}}

\caption{11}{Demonstration that Bob sees a train
moving toward him with speed $v$ as having a length that exceeds
its proper length by the relativistic Doppler factor $\sqrt{{1+v
\over 1-v}}$.}

\endinsert

The distance between these two events in Bob's frame (which is the
length he sees the train as having) is the distance between events
represented by the lower white circle and the black circle on Bob's
equiloc at its intersection with his equitemp through the lower white
circle.  That distance squared is the interval between the events, and
is therefore given by the area (in units of $\Omega_0$) of the larger
light rectangle, reproduced in the inset on the lower right.  The
squared proper length of the train, on the other hand, is given by the
area of the smaller light rectangle, reproduced in the inset on the
upper left.

Since the two rectangles have a side in common, the ratio of their
areas is just the ratio, ${1+v \over 1-v}$, of their aspect ratios, so
the length of the train as seen by Bob exceeds its proper length by
$\sqrt{{1+v \over 1-v}}.$ A similar construction establishes that when
the train is moving away, it is seen to be shorter than its proper
length by the factor $\sqrt{ {1-v \over 1+v}}$.

\bigskip

\head{9.  More spatial dimensions.}

The simplicity of these constructions diminishes as one introduces
additional spatial dimensions, but one feature of the higher
dimensional diagrams should be translated into the terminology
developed above.  Suppose Alice represents events in two spatial
dimensions in a three-dimensional space-time diagram, of which her
two-dimensional diagram is now one of a family of parallel
two-dimensional slices.  She applies the rules enunciated above to
every such two-dimensional slice.  She then aligns the slices so that
the one-dimensional equitemps in each slice associated with the same
time all lie in an equ\-itemporal plane perpendicular to the slices, and
so that the one-dimensional equilocs associated with the same position
in the first dimension all lie in a plane perpendicular to the plane
of the slices.  Each such plane of equilocs can be further resolved
into linear equilocs according to the positions they represent in the
direction orthogonal to the first spatial dimension.

The only remaining freedom lies in a new scale factor $\sigma$ relating
separation of equilocs in the same plane perpendicular to the slices,
to separation in space of the events they represent, in the direction
orthogonal to the first spatial dimension.  This is determined by a
requirement of isotropy: the two orthogonal photon trajectories
passing through any event in the 1+1 dimensional diagrams should
expand into a complete right circular cone of photon trajectories in
the 2+1 dimensional diagram.  Figure 12 demonstrates that this
determines the scale factor $\sigma$ in the perpendicular direction to
be the frame-independent geometric mean of the scale factors used in
the two-dimensional slices: $$\sigma =
\sqrt{\lambda\mu}.\eq(orthoscale)$$

\midinsert
\epsfxsize=5.5 truein
\centerline{\epsfbox{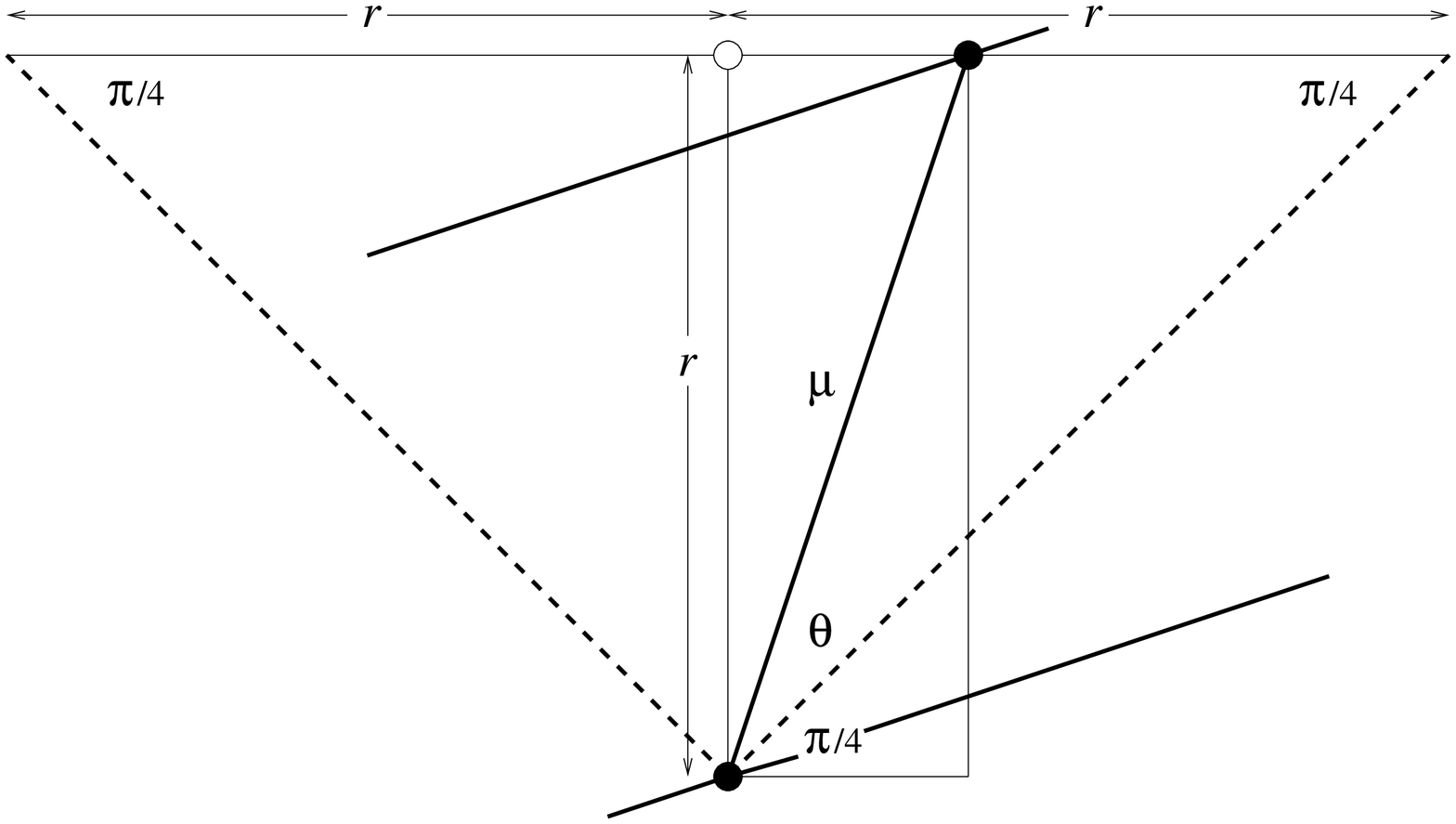}}

\endinsert

\caption{12}{A two dimensional slice of Alice's
(2+1)-dimensional space-time diagram.  Alice's equitemps are now
planes perpendicular to the slice that intersect it in the two
parallel heavy lines.  The third heavy line is an equiloc of Alice,
lying in the slice.  The two photon trajectories are the intersection
with the slice of a right circular cone of photon trajectories with
the lower black circle for its vertex, and the line connecting that
circle to the white circle for its axis.}

The figure shows a 2-dimensional diagram of Alice, now to be viewed as
a 2-dimension\-al slice of her 3-dimensional diagram.  Two of her
equitemps in the 3-dimensional diagram are parallel planes,
perpendicular to the 2-dimensional slice, that intersect it in the two
parallel heavy lines.  The two equitemps represent events 1 ns apart.
The third heavy line, connecting the two black circles is an equiloc of
Alice (that remains just a line in the higher dimensional diagram).
The black circles represent events 1 ns apart in Alice's frame.  The
dashed photon trajectories are the intersection of a right circular
cone of photon trajectories with the 2-dimensional diagram.  The axis
of the cone is the light vertical line connecting the lower black circle
with the upper white circle.

Consider a photon that connects the event represented by the black circle
on the lower equitemporal plane with an event on the upper
equitemporal plane that is displaced from the upper black circle in the
direction perpendicular to the 2-dimensional slice.  Since the
projection of the photon's trajectory into the plane of the slice
coincides with an equiloc in that plane, the motion of the photon out
of the plane represents its entire change in position in the
3-dimen\-sional diagram.  Since the two equitemporal planes represent
events 1 ns apart, the photon must cover a distance of 1 f, and it must
therefore finish displaced by the distance $\sigma$ from the upper black
circle in the direction perpendicular to the slice.

This final position must lie on the cone of photon trajectories
through the lower black circle, and must therefore be a distance $r$
from the white circle in the 3-dimensional diagram.  It is clear from
the diagram that $r$ is related to $\mu$ and $\theta$ by $$r =
\mu\sin(\pi/4+\theta).\eq(r1)$$ On the other hand the final position
of the photon is displaced from the upper black circle by a distance
$\sigma$ perpendicular to the slice, and displaced from the white
circle in the plane of the slice by a distance $\mu
\cos(\pi/4+\theta)$, so its distance in the 3-dimensional diagram from
the white circle must be $$ r = \sqrt{\sigma^2 +
\mu^2\cos^2(\pi/4+\theta)}.\eq(r2)$$ This is equal to the distance in
\(r1) provided $$\sigma^2 =
\mu^2\sin^2(\pi/4+\theta)-\mu^2\cos^2(\pi/4+\theta) =
\mu^2\bigl(-\cos(\h\pi+2\theta)\bigr) = \mu^2\sin2\theta =
\mu\lambda.\eq(done)$$ So the distance perpendicular to the
2-dimensional slice between equilocs containing events 1 f apart is
just the invariant quantity $\sqrt{\mu\la}$.  All observers with
equilocs in the plane of Alice's 2-dimensional slice use this same
scale factor for spatial separations perpendicular to that slice.
(This may well be the most difficult derivation of $y'=y, z'=z$ ever
to appear in the literature.)

\head{10. Related work.}

This way of developing 1+1 dimensional flat space-time diagrams
refines and extends an approach I recommended several years ago for
teaching special relativity to nonscientists who know a little
elementary algebra and plane geometry [2,3].  An expanded exposition
can be found in my forthcoming book [4] on special relativity for
nonscientists.  Dieter Brill and Ted Jacobson have recently given a
similar geometric treatment of the interval [5].  I learned from Brill
and Jacobson that features of this point of view go back at least to
1913 [6], and that some beautiful film clips illustrating some of
these geometric relations are at the website  of Dierck-Ekkehard
Liebscher [7]. Closely related material can be found in Liebscher's
book [8].

\medskip

\head{Acknowledgment.}  

Supported by the U.S.~National Science Foundation, Grant No.~0098429.
\bigskip

\noindent {\bf References}
\medskip

\item{[1]} Robert Geroch, {\it General Relativity from A to B\/},
University of Chicago Press, 1978.

\item{[2]} N. David
Mermin, ``An Introduction to Space-Time Diagrams'', American Journal
of Physics {\bf 65}, 476-486 (1997). 

\item{[3]} N. David Mermin, ``Space-time intervals as light rectangles'',
American Journal of Phys\-ics {\bf 66}, 1077-1080 (1998).

\item{[4]} N. David Mermin, {\it It's About Time: Understanding
Einstein's Relativity\/}, Princeton University Press, to be published in
2005.

\item{[5]} Dieter Brill and Ted Jacobson, ``Spacetime and Euclidean
Geometry'', {\tt http://arxiv. org/abs/gr-qc/0407022}.

\item{[6]} Edwin B.~Wilson and Gilbert N. Lewis, ``The space-time
manifold of relativity.  The non-euclidean geometry of mechanics and
electromagnetics'', Proc. Amer. Acad. Boston {\bf 48}, 389-507 (1913).

\item{[7]} {\tt http://www.aip.de/\hskip -3pt $~\tilde{\ }$lie/}

\item{[8]} D.-E. Liebscher, {\it Einsteins Relativit\"atstheorie und
die Geometrien der Ebene\/}, Teubner Stuttgart (1999).

\bye

\vfil\eject

\midinsert
\epsfxsize=6 truein
\centerline{\epsfbox{ann-fig-1.eps}}
\endinsert

\vfil
\centerline{\bf Figure 1}

\eject

\midinsert
\epsfxsize=5truein
\centerline{\epsfbox{ann-fig-2.eps}}
\endinsert

\vfil
\centerline{\bf Figure 2}

\eject

\midinsert
\epsfxsize=4truein
\centerline{\epsfbox{ann-fig-3.eps}}
\endinsert

\vfil
\centerline{\bf Figure 3}

\eject

\midinsert
\epsfxsize=5.5truein
\centerline{\epsfbox{ann-fig-4.eps}}
\endinsert

\vfil
\centerline {\bf Figure 4.}

\eject

\midinsert
\epsfxsize=5truein
\centerline{\epsfbox{ann-fig-5.eps}}
\endinsert

\vfil
\centerline{\bf Figure 5.}

\eject

\midinsert
\epsfxsize=5.5truein
\centerline{\epsfbox{ann-fig-6.eps}}
\endinsert

\vfil
\centerline{\bf Figure 6.}

\eject

\midinsert
\epsfxsize=7truein
\centerline{\epsfbox{ann-fig-7.eps}}
\endinsert

\vfil
\centerline{\bf Figure 7.}

\eject

\midinsert
\epsfxsize=5truein
\centerline{\epsfbox{ann-fig-8.eps}}
\endinsert

\vfil
\centerline{\bf Figure 8.}

\eject

\midinsert
\epsfxsize=6truein
\centerline{\epsfbox{ann-fig-9.eps}}
\endinsert

\vfil
\centerline{\bf Figure 9.}

\eject

\midinsert

\epsfxsize= 6truein
\centerline{\epsfbox{ann-fig-10.eps}}
\endinsert

\vfil
\centerline{\bf Figure 10.}

\eject

\midinsert
\epsfxsize= 7truein
\centerline{\epsfbox{ann-fig-11.eps}}
\endinsert

\vfil
\centerline{\bf Figure 11.}

\eject

\midinsert

\epsfxsize= 7 truein
\centerline{\epsfbox{ann-fig-12.eps}}
\endinsert

\vfil
\centerline{\bf Figure 12.} 

\eject
\end